\documentclass[letterpaper, 11pt]{article}
\pdfoutput = 1

\usepackage{shortcuts}

\usepackage[margin = 2.7cm]{geometry}
\usepackage[nottoc]{tocbibind}

\makeatletter
\renewenvironment{abstract}{%
    \if@twocolumn
      \section*{\abstractname}%
    \else 
      \begin{center}%
        {\bfseries \large\abstractname\vspace{\z@}}
      \end{center}%
      \quotation
    \fi}
    {\if@twocolumn\else\endquotation\fi}
\makeatother

\begin{document}

\thispagestyle{empty}

\begin{flushright}
\texttt{BRX-TH-6706}
\end{flushright}

\vspace*{0.1\textheight}

\begin{center}

\textbf{\LARGE Truth, or mere Beauty?}\\ \smallskip
{\Large Simple thoughts on nearly everything}

\vspace{0.06\textheight}

{\large \DJ or\dj e Radi\v cevi\'c}
\vspace{0.02\textheight}

{\it Martin Fisher School of Physics\\ Brandeis University, Waltham, MA 02453, USA}\\

\vspace{0.12\textheight}

\begin{abstract}
\normalsize
High energy physics features many ingenious tools for extracting finite results from formally divergent expressions. This brief note argues from a new perspective that all such formal infinities are meaningful markers of new physics. As such, they deserve to be explored in detail --- and even the simplest quantum field theories have more of them than is commonly thought.
\end{abstract}

\end{center}

%
%

\newpage

\section{Introduction}



Many fundamental questions about our world remain unresolved despite the tremendous progress seen in high energy physics over the last century. What other particles exist beyond the Standard Model?  What sets the huge hierarchies between scales that appear in nature?  How does gravity (and in particular a universe expanding at an accelerating rate) consistently fit into the framework of quantum field theory? And how are all these theories actually defined --- or, in other words, what are the underlying fundamental degrees of freedom and their dynamical laws?

Questions such as these are deep, difficult, and possibly even ill posed. There is no guarantee that anyone alive in 2022 will get to see them resolved. Conversely, any pronouncement of a ``crisis'' in high energy physics that is based on the lack of such fundamental answers is more of a product of the pronouncer's political and psychological predilections. (While the field as a social endeavor might implode if enough members deem it to be in a crisis, this would still not objectively show how off the mark its current theories might have been.) Simply put, it is quite difficult to tell how close this discipline is to achieving its primary objective. It would be nice to have more hints about the state of our knowledge.

The purpose of this essay is to point out one precise way in which every currently popular approach to high energy physics \emph{fails} to fully describe the fundamental structure of the universe. This is not meant to be some sort of a dour dissection of a diseased discipline. The goal here is simply to articulate an interesting and underappreciated tension that, on its own, might never become a research topic during the remaining lifespan of this scientific field.

Despite what one might suspect, this is also not a \emph{cri de c\oe ur} for more mathematical rigor in physics. The issue at hand actually originates in pure math. It arguably goes back to the old Pythagorean claim that the diagonal of the unit square, i.e.\ the number $\sqrt 2$, is ``unutterable'' \cite{Kramer:1983}. In modern (though not completely standardized) parlance, the tension that will be described here is associated to what may be called the \emph{finitist/formalist split} in the foundations of mathematics. Versions of this dichotomy have been debated many times; some relevant references are e.g.\ \cite{Brouwer:1908, Heijenoort:1967, Snapper:1979, Bishop:1985, Martin-Lof:2008, Lenchner:2020}. Now it is time to discuss the natural avatar of this split in high energy physics.

To start, consider the following reasonable-sounding proposition, the \emph{weak finitist axiom}:
\begin{center}
  \textbf{No measurement can yield an infinite result.}
\end{center}
Informally, this means that every measuring device has a resolution (the minimal difference between numbers it can output) and a range (the maximal and minimal outputs). As detailed below, even this proscription of infinities will turn out to have far-reaching consequences. 

Instead of talking about measurements, which are rather loaded concepts in quantum theory, one can subsume the set of ``measuring devices'' into the broader class of ``physical systems.'' This leads to the main focus of the present discussion, the \emph{strong finitist axiom}:
\begin{center}
  \textbf{Every physical system must be fully defined without any infinities.}
\end{center}
Although information-theoretic ideas will not be used here, the strong finitist axiom can also be understood to say that no physical system can contain an infinite amount of information.

\newpage

One very instructive consequence of either finitist axiom is that no measurement can return infinitely many digits at once. This in particular implies that $\sqrt 2$ cannot be the output of any realistic experiment, as Pythagoras may have intuited. From a practical standpoint, this is actually a perfectly reasonable position. Imagine drawing a unit square and measuring its diagonal with ever-increasing precision. It will eventually turn out that the notion of length cannot be resolved more finely than about a nanometer --- the scale at which it becomes possible to discern individual atoms of the ink used to draw the square. Ultimately, this limitation holds because any line one can draw must consist of finitely many finitely-sized particles of ink.

Of course, this does not mean that $\sqrt 2$ cannot be defined. It can be calculated to arbitrary precision using an expansion like
\bel{\notag
  \sqrt 2 = \sum_{n = 0}^\infty \binom{2n}{n} \frac1{8^n},
}
and it can be arithmetically manipulated in all the usual ways. But its full calculation is always possible only \emph{in principle}: although the algorithm itself can be described succinctly, and although one can prove theorems about arbitrarily many digits of $\sqrt 2$, it would still take an infinite amount of time to actually calculate all these digits, and it would take an infinite amount of memory to store them. No one has such resources on hand\ldots\ Right? 

The \emph{formalist axiom} in physics is the claim that ``infinite information'' systems, such as those that can encode all digits of $\sqrt 2$, are in every sense as physical as the ``finite information'' ones. For simplicity, the former systems will be called \emph{formal}, and the latter ones will be called \emph{finitary}. Any physical model with an infinite number of internal states or degrees of freedom would thus be formal. The simple harmonic oscillator (SHO) with its infinite ladder of energy eigenstates is a very familiar example of such a formal system. (It is easy to construct a superposition of infinitely many SHO eigenstates that encodes all the digits of $\sqrt 2$.) Textbook versions of quantum field theory and general relativity are formalist from the get go, relying as they do on SHOs, differential geometry, and classical Lagrangian mechanics constructed from real numbers. The same holds for most of modern high energy research, where the formalist axiom seems so natural that it is not given a second (or even a first) thought.

The advertised tension should by now be apparent. If no physical system can contain an infinite amount of information, then \emph{all formal models must be wrong descriptions of our world}. In other words, if the finitist axiom is true, the formalist one is not, and current high energy theories must fail to fully capture the fundamental dynamics of our world. This does not mean that formal systems cannot be excellent approximations of the world, but it does mean that they \emph{must} break down when probed sufficiently deeply.

Two questions now naturally present themselves:
\begin{enumerate}
  \item Is the finitist axiom actually true?
  \item If so, what are the implications for modern (formal) theories?
\end{enumerate}
The remainder of this text will be dedicated to answering these questions.
Section \ref{sec finitism} will answer the first question in the affirmative using a new and very general argument. This is likely the most important novelty introduced in this note. Section \ref{sec formalism} will answer the second question by presenting a broad illustration of the kind of finitist structure that must be introduced to the existing formal models to give them a chance of being true descriptions of reality.

\newpage

\section{Finitism is true} \label{sec finitism}

The finitist axioms presented above were motivated by the idea that no physical apparatus can store or process an infinite amount of data. This is certainly \emph{truthy}, but is it actually \emph{true}?  What if our experimental protocols are imperfect, but the underlying physical degrees of freedom still somehow encode and evolve an infinite amount of information in finite time and space? For example, what if the basic building blocks really are just coupled SHOs, and we merely lack the smarts to efficiently probe the physical processes that involve infinitely excited states?

There exists a standard set of arguments against these formalist scenarios (and thus in favor of finitism). They all involve the fact that gravity exists. Gravity implies that the amount of information that can be stored in any finite amount of space must be bounded; exceeding this \emph{Bekenstein bound} must result in a black hole formation, contradicting what we see \cite{Bousso:2002ju}.

Invoking such formidable argumentation feels a bit like using a sledgehammer to crack a nut --- except, in this metaphor, we also do not even know how to safely operate a sledgehammer. We understand gravity at short distances worse than we understand usual quantum field theory. Within (quantum) gravity, we understand black holes much worse than nonsingular spacetimes. Gravitational arguments should     tilt our priors towards believing the finitist axiom, but they are still complicated enough that it is hard to accept them as proof that nature is finitary.

Fortunately, there also exists a much more elementary argument in favor of finitism. It applies to all quantum theories, even those that do not feature gravity. It is also rather subtle and may at first appear quite controversial.

The key observation is suspiciously simple: \emph{there are no infinities in standard mathematics}. Any rigorous definition of an infinity ultimately regularizes it --- that is, replaces it with a large but finite number, and then asks that all other numbers one considers stay much smaller than this regulator. The distinction between finite and infinite (or infinitesimal) quantities then simply becomes the distinction between quantities that do not depend on the regulator and those that do. As a consequence, any formal physical system with an infinite number of degrees of freedom is guaranteed to turn into a finitary system upon sufficiently close inspection. In this sense, there is no such thing as a truly infinite system, the strong finitist axiom is true practically by definition, and the various references to ``infinity'' on the preceding pages were merely sloppy verbiage.

This attitude may seem counterintuitive, wrong, or even deeply sacrilegious. To make it more palatable, here are four examples that show how a formally infinite (or infinitesimal) quantity becomes replaced by a \emph{detectably} finite one when the rigorous definition is taken seriously.

\bt{Example 1:} One way to define the derivative of a function $f$ at point $x$ is to pick an $\eps > 0$ and look at the ratio
$$
  f'_\eps(x) \equiv \frac{f(x + \eps) - f(x)}{\eps}.
$$
This is defined whenever $f(x)$ and $f(x + \eps)$ are.  However, the derivative is only said to \emph{exist} if $\eps$ can be chosen so small that $f'_\eps(x) = O(\eps^0)$, i.e.\ so that $|f'_\eps(x)|$ is much smaller than any large scale set by $\eps$. Then one can take the $\eps$-independent piece $f'(x)$ and call that \emph{the} derivative at $x$.

Discarding these $\eps$-dependent terms is a formal step. It makes sense because ``$\eps$ is very small,'' but it cannot be done with complete impunity. To see why, consider higher derivatives. For example, the $\eps$-independent part of the first derivative of $f(x) = x^2$ is zero at both $x = 0$ and $x = \eps$, but the $\eps$-dependent parts are needed to show that the second derivative is $f''(0) = 2$.

The situation becomes even more delicate when working with derivatives of very high orders. In this case the putatively small terms that would naively be thrown away while extracting the $\eps$-independent piece might turn out to actually be finite. For instance, take $f(x) = \e^x$, which is expected to have $f^{(n)}(0) = 1$ for every $n$. But if the $n$'th derivative is actually defined via $f^{(n)}_\eps(x) \equiv \big[f^{(n - 1)}_\eps\big]'_\eps(x)$ for some fixed $\eps$, say $\eps = 1/M$ for a large $M \in \mathbb N$, then a quick calculation shows that the $O(\eps)$ term in $f^{(M + 1)}_\eps(0)$ is actually proportional to $\eps M = 1$. In other words, if $f^{(M + 1)}_\eps(0)$ is viewed as a polynomial in $\eps$, the $\eps$-linear term is comparable to the $\eps$-independent term, which is equal to unity as per the standard expectation that $f^{(M + 1)}(0) = 1$.

The issue here is that one has to commit to a fixed value of $\eps$ when defining the derivative --- and no matter how small this value is, it is \emph{always} possible to look at a derivative of order $n > 1/\eps$ and find that terms linear in $\eps$ are actually comparable in size to $\eps$-independent terms. While $\eps$ can be chosen small enough to lead to a good definition of the first derivative, one cannot ``move the goal posts'' and decrease $\eps$ depending on the order of derivative one wishes to calculate. If one wishes to define an arbitrarily high derivative, the only sufficiently small $\eps$ would then be zero, which would be meaningless. The upshot is that $\eps$ must be a small but finite quantity that cannot be divorced from the definition of a derivative --- and that can be detected.

\bt{Example 2:} Like $\sqrt 2$, the number $\e$ is an example of an ``infinite information'' quantity that is only physical within a formalist point of view. It can be defined as
$$
  \e = \sum_{n = 0}^\infty \frac 1{n!}.
$$
The standard way to give rigorous meaning to any infinite sum is to replace it by a partial sum, in this case
$$
  \e_M \equiv \sum_{n = 0}^M \frac 1{n!}
$$
for some large integer $M$. The regularized quantity $\e_M$ depends on such an $M$ very mildly, as it satisfies $|\e_{M + 1} - \e_M| = \frac 1{(M + 1)!} \ll 1$. This is similar to the behavior of the regularized derivative $f'_\eps(x)$ from the previous example: as long as the regulator (whether $M$ or $1/\eps$) is large enough, it practically does not matter what value it takes, as there is an underlying regulator-independent quantity that one can imagine to be working with in lieu of the regulator-dependent one. In other words, one can imagine formally dropping small, $M$-dependent terms from $\e_M$ to obtain $\e$.

As before, however, one may probe the cutoff used in the rigorous definition of the number $\e$. A straightforward way to do so is to consider very high powers of $\e_M$, which are given by
$$
  (\e_M)^t
   \approx
  \e^{\,t\, [\,1\, -\, O(\, 1/(M+1)!\, )\,]}.
$$
For $t \sim (M + 1)!$, the value of $(\e_M)^t$ will be appreciably different from the ``target value'' $\e^t$, which can be defined as the $M$-independent part of the regulated quantity $(\e^t)_M \equiv \sum_{n = 0}^M \frac{t^n}{n!}$. More precisely, for any $M$ that one might use in the rigorous definition $\e_M$ of $\e$, there exists a sufficiently high power $(\e_M)^t$ that strongly deviates from the cutoff-independent part of $(\e^t)_M$ that one would like to use to define $\e^t$. There is no running away from the $M$-dependence.

Here is a fun idea. The time dependence of quantum variables often looks like $\O(t) = \e^{\i E t} \O$. The purely mathematical result above now says that, if you wait long enough, quantum mechanics will discover corrections to the number $\e$. This ludicrous notion will reappear in the next Section, where it will be treated with a little bit more care.

\newpage

The last two examples can be rephrased using the familiar notion of a limit. The regulator-dependence seen in both examples can be understood as a statement that two limits do not commute. For instance, the order of limits $t \rar \infty$ and $M \rar \infty$ turns out to matter when calculating $(\e_M)^t$. Indeed, the claim of this paper is precisely that it is always possible to construct some limiting procedure (such as looking at high powers or high orders of derivatives) in which the noncommutation of limits allows the initial regulator to be detected.

\bt{Example 3:} Sometimes, the regulator is left rather implicit, and it is not immediately clear what limiting procedure is used to define a mathematical concept. For example, one may define the set of real numbers $\R$ not in terms of decimal expansions or infinite series, but rather more abstractly as a one-dimensional Euclidean manifold. Furthermore, this manifold need not be defined following Euclid's axioms, but as a particular kind of topological space equipped with an appropriate metric. What regulators are in play here?

Here is a sketch of an answer. A set is promoted to a topological space by introducing a notion of ``nearness,'' i.e.\ by specifying which of its elements can be considered to be ``neighbors.'' This is standardly done as follows. Given a set $\bb X$, a topological space $(\bb X, \cal T)$ is defined by picking a collection $\cal T$ of subsets of $\bb X$ (the so-called ``open sets'') that satisfy the following properties:
\begin{enumerate}
  \item The set $\bb X$ and the empty set both belong to $\cal T$ (i.e.\ are open sets),
  \item Any union of open sets is an open set,
  \item Any intersection of a finite number of open sets is an open set.
\end{enumerate}
There are two large parameters that can enter this construction. One is the cardinality $N$ of $\bb X$. The other is the maximal number $M$ of open sets whose intersection is demanded to still be open. Both of these numbers are allowed to be large (i.e.\ one can consider the limit $M, N \rar \infty$), but the third line in the definition requires that at all times one keeps $M$ much smaller than $N$ or \emph{any other} large scale.

When interpreting $\R$ as such a topological space, $N$ is simply the regularized cardinality $|\R|$. The meaning of $M$ is subtler. Roughly, it sets the maximal resolution with which $\R$ is probed by open sets --- an analogue to the number of digits one keeps in a decimal expansion. It is a purely ``ultraviolet'' regulator. This is in contrast to $N$, which counts how many different numbers $\R$ contains. That is, $N$ includes both the ``ultraviolet'' and ``infrared'' information. The large ratio $N/M$ can (again, roughly) be interpreted as the largest number included in $\R$ once it is regularized, i.e.\ as the ``infrared'' regulator of $\R$.

A rigorous definition of a topological space over the set of real numbers thus constructs a regularized topological space $\R_{N, M}$ for $N \gg M \gg 1$. (This space can then be straightforwardly equipped with a metric, this metric can be used to define limits, and so on.) The presence of regulators means that any real function can be ultimately viewed as a homeomorphism between finite but large topological spaces. This means that any large scales used in defining a real function will ultimately be expressed via $N$ and $M$. This can provide another perspective on the issue of defining derivatives from Example 1.

Now that it is clearer which regulators are in play, it is also possible to think of ``experiments'' that probe them. The ``ultraviolet'' regulator $M$ can be probed by considering high powers or high orders of derivatives of real functions. The ``infrared'' regulator $N/M$ is probed more prosaically, by simply taking the argument of an unbounded function like $f(x) = x^2$ to large values.


\newpage

\bt{Example 4:} Having ploughed through the previous examples, an inquiring mind might now want to know whether the set of natural numbers $\bb N = \{1, 2, 3, \ldots\}$ also needs regularization. Natural numbers can be constructed inductively from purely ``finite information'' quantities: start from $1$, add $1$ to get $1 + 1 = 2$, then add $1$ again to get $2 + 1 = 3$, and so on. Surely $\bb N$ can then be defined as ``the set of all numbers that can be obtained by this procedure,'' thereby giving an infinite set that never got regularized along the way?

This is a dramatic example of how human intuition can clash with mathematical rigor. The phrase ``and so on'' from the last paragraph seems quite clear, but that does not mean that it has a ready mathematical meaning. In fact, there is no reason why a natural English phrase should correspond to a meaningful mathematical object, as (for example) illustrated by the incompleteness theorems of G\"odel (himself a prominent doubter of formalism in math). In the present case, ``and so on'' \emph{can} be rigorously defined, but doing so introduces a regulator. Indeed, this inductive definition of $\bb N$ simply states that
$$
  \bb N = \bigcup_{n = 1}^\infty \{n\}.
$$
Just like in Example 2, the infinite union is really defined as a union of $M \gg 1$ sets. Similarly, the cardinality of $\bb N$ is
$$
  |\bb N| = \sum_{n = 1}^\infty 1,
$$
which is again rigorously approached (and shown to diverge) by introducing a large cutoff $M$.

It is incredibly important to stress that this argument in no way claims that the $M \rar \infty$ limit of $\bb N_M \equiv \bigcup_{n = 1}^M \{n\}$ is meaningless. Indeed, this is a perfectly fair object of study, and it is replete with all sorts of wonderful properties like Fermat's last theorem. The present discussion merely aims to point out that all number-theoretic theorems are rigorously stated using the set $\bb N_M$ with the understanding that the limit $M \rar \infty$ is taken first, so that no matter what number $K \gg 1$ one later works with, the set $\bb N_M$ is always guaranteed to have numbers much greater than $K$. In other words, whenever one might want to define a mathematical quantity using a limit $K \rar \infty$, one promises to still take $M \rar \infty$ first. This is certainly meaningful.

So much for examples. Two points now deserve particular emphasis:
\begin{itemize}
    \item At this stage, it is not at all clear whether physical theories should be barred from e.g.\ taking the time $t$ to be much larger than every other quantity in play, including potential regulators for the set $\bb N$ or the number $\e$. It may also be foolhardy to hope that such regulators will be physically detectable in the foreseeable future. A reasonable approach would be to keep an open mind and study all possible orders of limits that can be applied to regulators. Ultimately, the goal of physics should be to devise \emph{experiments} that will \emph{measure} which order of limits is realized in nature --- and coming up with such experiments will be very difficult if the whole field commits to a fixed order of limits from the outset!

    \item None of this \emph{proves} either finitist axiom. It is, unfortunately, impossible to discuss every mathematical definition and pinpoint where the large parameters lie, the way it was done here for derivatives, the number $\e$, topological spaces, and the set of natural numbers. These four examples hopefully help illustrate the general point: whenever one is used to thinking of infinite quantities, a rigorous examination shows the existence of a regulator. The readers (if any) are invited to repeat these exercises for their favorite mathematical concepts.
\end{itemize}
\newpage

\section{Formalism is beautiful, but\ldots} \label{sec formalism}

Regulators are icky. They complicate formul\ae\ while barely affecting the answers to most questions of practical use. It is no surprise that some of the most important technical tools in theoretical physics are those that push explicit regulators further under the rug. Examples include calculus, the central limit theorem, renormalization and effective field theory, and even category theory. Each of these gives a streamlined framework for calculations in which it is seldom necessary to keep track of all the $\eps$'s and $\delta$'s. The fewer regulators, the more beautiful the framework.\footnote{
  The similarity between this attitude and certain political philosophies is perhaps worth thinking about.
}

The previous Section has argued that no matter how beautiful a formal system may seem, the existence of regulators becomes important when one starts asking more fundamental questions. Examples of four such questions were discussed there. Other questions may be more complex and lead to extensive debate. Why can a divergent sum like $1 + 2 + 3 + \ldots$ be replaced by $-\frac1{12}$, and what other sums can be treated this way? When will the law of large numbers fail and what governs the distribution of ``rare events?'' What kinds of corrections to an effective field theory must be observed at high temperatures, densities, or curvatures?  How about the phenomena at late times and/or spatial infinity, such as black hole evaporation?

Instead of discussing such involved examples in the same manner as before, this Section will merely set the stage for similar discussions through a broad overview of the issue of regulators in quantum field theory (QFT). Now, QFT is a vast subject that can include everything from single-qubit quantum mechanics to classical field theories (such as general relativity) to string theory. Consequently, QFT may have many infinities to be regularized, and some are much more commonly discussed than others. The goal here is to systematize this zoo of regulators and to show that there exist \emph{four} types of regularization that must be distinguished in a generic QFT.

A useful way to start this analysis may be by asking whether there is a QFT that requires no regularization whatsoever. Most QFT textbooks start by studying a free scalar field theory, but despite its simplicity this is most certainly not a theory free of infinities. (Its pedagogical value lies in the fact that you can ``shut up and calculate'' without seriously worrying about infinities.) The infinities that \emph{are} brought up in this context are usually associated to short distances in position space, and they can be probed by turning on various relevant interactions.

Even putting a scalar field theory on a finite spatial lattice (i.e.\ regularizing the most commonly studied infinities) does not remove all formal traits from the theory. The target space of such a lattice theory remains infinite --- the theory is literally a collection of coupled SHOs that was discussed in the Introduction as an example of formalism in physics.

What happens if the target space is also regularized? Or, to simplify things further, take a theory in which the target space is far from infinite: a system of spins, or qubits, living on sites of a spatial lattice. Here the Hilbert space has a manifestly finite dimension, $D = 2^N$, where $N$ is the number of lattice sites. Is this theory free of implicit infinities?

The answer is \emph{no}. One reason is that this is still a standard quantum mechanical system which can evolve for an arbitrarily long time. One can also, at least in principle, expose such a system to repeated perturbations with an infinitesimally small time separation. In other words, the rules of quantum mechanics still allow time to be a real number, and the previous Section has already argued that $\R$ cannot be defined without introducing further regulators. Another way to say this is that in the path integral language the temporal direction still needs to be latticized.

Finally, what happens if time gets regularized, and the evolution of a given system with a finite-dimensional Hilbert space is now a discrete process, like a quantum circuit?  Is this setup free of infinities yet?

The answer is still \emph{no}. There is one subtle source of infinities left. The Hilbert space is still a vector space over the field of complex numbers $\C$, which is just as much in need of regularization as $\R$ was. Time evolution, even if it proceeds in discrete steps using only a finite set of quantum gates, is still effected by multiplying generic $D$-dimensional vectors by $D \times D$ unitary matrices. To regularize the infinity of possible state vectors it is necessary to go beyond the established quantum postulates and to build a formalism in which the quantum states live in a vector space over a \emph{finite} field. (Alternatively, the operator algebra must be taken to be a finite subalgebra of the algebra $\C^{D \times D}$ of complex matrices.) This is a rather dramatic change to the rules of quantum mechanics. Its ramifications are by and large unknown.

This study of scalar field theory in fact highlights all four different places where regulators are (implicitly or explicitly) used when talking about a generic QFT. To recap, these are:
\begin{enumerate}
  \item The space of scalar numbers over which the Hilbert space is defined,
  \item The range of the time parameter,
  \item The target space (i.e.\ the set of basis states for each degree of freedom),
  \item The position or momentum space (i.e.\ the set of all degrees of freedom).
\end{enumerate}
The first two sources of infinities are present in every quantum theory, be it a single Ising spin or M-theory. Infinities of the third kind exist in any theory where the target space is not a finite set; this includes any theory of bosonic fields, from ordinary scalars to $\sigma$-models and Yang-Mills gauge theories. The fourth is present in any theory with many locally coupled degrees of freedom, i.e.~whenever the QFT is in more than $0 + 1$ spacetime dimensions. In path integral/statistical mechanical formulations of QFT, the second and fourth sources of infinities are usually viewed as one, but their difference is stark in a canonical/Hamiltonian framework.

As already mentioned, not all infinities are equally important in practice.  Correlation functions at very small temporal or spatial separations probe infinities from the second and fourth category, and it is fair to say that their study has historically driven most work on QFT \cite{Wilson:1971bg}.

Infinities from the third category are well studied in the context of quantum mechanics, where they feature in rigorous constructions of rigged Hilbert spaces and other similar axiomatizations \cite{Gelfand:1964, Reed:1972}. Such issues are less studied in higher-dimensional QFT. One reason is that in the most common QFTs one simply does not consider observables that probe these infinities. Nevertheless, some target space infinities may still enter the mainstream through a back door. For instance, if the target space is a manifold of a very large dimension --- say $\SU(N)$ at $N \gg 1$ --- the parameter $N$ will turn out to be crucial in describing the behavior of the resulting theory \cite{tHooft:1973alw}.

Infinities from the first category, while in a sense the most basic, are practically unheard of. An interesting recent claim is that they are connected with the measurement problem in quantum mechanics. Very roughly, the idea is that the data hidden by a cutoff on the space of complex numbers essentially act as hidden variables: as time passes, the cutoff moves, and the hitherto hidden digits determine how the wavefunction will collapse in each given instant \cite{Gisin:2018ihz, Gisin:2019emq}. These ideas will not be discussed here but they deserve further scrutiny.

Recall that the definition of $\R$ as an infinite topological space required \emph{two} regulators, an ``infrared'' and an ``ultraviolet'' one. Each category of infinities described above may similarly require more than one regulator. One way to systematically keep track of all the needed regulators is by thinking of each of the four sources of infinity as a manifold that needs to be regularized.

Formally, the space of scalar numbers is $\C$, and the range of the time parameter is $\R$ (or $S^1$ if working in Euclidean time at finite temperature). These are flat manifolds. In general, $d$-dimensional flat manifolds can be regularized very simply, by viewing them as huge hypercubic lattices, possibly with some specific boundary conditions. The size of the lattice (i.e.\ the number of sites) in each direction is one set of regulators needed here. However, as discussed above, specifying the regularized size is not enough to capture the structure of a manifold as a topological space. A second set of regulators is needed to specify how many sites can be considered to be ``in the neighborhood'' of any given site along each of the $d$ directions. Thus for a flat $d$-dimensional manifold one should expect to have at least $2d$ regulators. Of course, in practice it will typically be convenient to assume that the regulators along different dimensions are the same, but such assumptions must be questioned e.g.\ when performing dimensional reduction.

The target space and the position space are often much more complicated curved manifolds. Luckily, a key fact about manifolds is that they are locally Euclidean, i.e.\ they can be broken up into small pieces that look flat. A general $d$-dimensional manifold can thus be viewed as a patchwork of Euclidean $d$-dimensional manifolds, each with $2d$ regulators that govern its linear size and the size of minimal open sets or ``neighborhoods'' along each dimension. A new set of regulators must now enter, however, in order to govern the number of these Euclidean patches. This brings the total number of regulators to at least $3d$. As before, a typical application will impose certain symmetries and may well have only three different regulators in the end --- the linear size of the manifold, the linear size of a Euclidean patch, and the linear size of a ``neighborhood'' in each patch. On the other extreme, it is also possible to imagine perverse scenarios in which e.g.\ each Euclidean patch has some kind of gerrymandered shape whose dimensions are set by multiple scales that all differ patch to patch.

While this discussion is pretty abstract, these regulators do have intuitive physical meanings. The inverse of the number of lattice sites is typically referred to as the \emph{lattice spacing}, and this is usually the only regulator that is explicitly discussed when trying to build continuum QFTs out of finite ones. The size of a ``neighborhood'' in position space is the \emph{smoothing} (or \emph{smearing}) scale, as the degrees of freedom on nearby points cannot be readily distinguished from each other; indeed, in formal QFT, it is always assumed that an operator near a point $x$ can be expressed as $\O(x + \eps) \approx \O(x) + \eps \,\del \O(x)$ with the understanding that the second term on the r.h.s.\ is ``small.'' The smoothing scale is the regulator that controls this expansion. The size of Euclidean patches in the target space is what practitioners might call the \emph{size of quantum fluctuations} in the given field, and each different patch corresponds to a different \emph{background} around which the field fluctuates. Very explicit examples of all these scales in action can be seen by studying the continuum limits of lattices of quantum rotors \cite{Radicevic:2D}.

It is also possible to demand that the manifold have a lot more structure than what a basic topological space provides. For example, manifolds are often also fiber bundles, or sheaves, with nontrivial connections describing how the local patches are pieced together. This way one can discuss manifolds that are spin, symplectic, K\"ahler, and so on. Adding such extra structure typically increases the number of regulators in play. For example, rigorous definitions of U(1) connections will come with their own regularization.

Coming full circle, how does this systematization of regulators ultimately apply to that venerable textbook QFT, the scalar field in $d$ spatial dimensions? To be specific, this is a \emph{noncompact} scalar field, so its target space is $\R$. Its position space is $\R^d$, assuming no periodic boundary conditions. Thus all four classes of regulators come from flat manifolds. Each dimension of each manifold comes with two regulators, a lattice spacing and a smoothing scale. To simplify matters, assume that all spatial dimensions have the same regulators, and also that the space of complex numbers (the first category of infinities) is not regularized, so that the rules of quantum mechanics remain the familiar ones. This still gives a total of \emph{six} different regulators. These are the lattice spacings and smoothing scales in the spatial, temporal, and target directions.

There is no a priori reason why these regulators would enjoy any special relations between themselves. (Of course, the exception here is that the smoothing lengthscale must always be much larger than the corresponding lattice spacing.) This means that a generic noncompact scalar QFT, once carefully defined, will have at least six different regulators in play. Small wonder that no \ae sthetically minded practitioner wants to keep track of all this junk!

Things get more complicated when one introduces multiple species of scalars, or when one attempts to gauge a symmetry. The latter step requires splitting the manifold into local Euclidean patches, interpreting it as a base of a principal fiber bundle, and defining connections between the patches. As hinted already, this all comes with further regulators. Thus to rigorously define the Standard Model one may be expected to juggle $O(100)$ generally different cutoffs. At this point the insistence on formalism is no longer a question of keeping the framework beautiful, but rather of making sure it can be used at all by mere mortals.

This preponderance of regulators in the Standard Model has one important benefit though: accepting their existence alleviates any acute anguish arising out of assorted hierarchy problems. Instead of worrying about the huge difference between the QCD scale and the Planck scale, say, one may view this ``gap'' as a natural placeholder for some of the dozens of other scales that may appear as regulators of our formal theories. These scales are not experimentally associated to new particles, but rather to qualitative changes to the rules governing the particles we do see. It would be fascinating to further flesh out this fledgling flight of fancy.

Finally, now it makes sense to think more about the strange idea from Example 2 in Section \ref{sec finitism}. The time evolution of a standard quantum system with a $D$-dimensional Hilbert space is described by a $D\times D$ unitary matrix of the form $U(t) = \e^{\i H t}$ for any $t \in \R$. For any fixed Hamiltonian $H$, one can imagine waiting long enough to probe the cutoff used to rigorously define the number $\e$. One may also imagine that the rigorously defined domain of validity of the time parameter $t$ is so small that the cutoff used in $\e$ cannot be probed with the given $H$. However, if the Hamiltonian is simply rescaled by a sufficiently large number, $U(t)$ can once again become sensitive to the ``graininess'' of $\e$. In other words, a system with a large enough characteristic energy, if left to evolve for long enough, will eventually discover the precise, cutoff-dependent definition of the exponential function that describes quantum evolution. (The fact that $\e^{\i Ht}$ is periodic does not qualitatively alter this conclusion because the period itself is irrational, and hence waiting long enough would also detect the deviations of this period from the ``infinite information'' number $2\pi$.) Of course, perhaps the preceding sentences may turn out moot because the exponential function in $U(t)$ must be replaced by something else when recasting the rules of quantum mechanics into a fully finitary framework. Either way, at least some regulators used to construct $U(t)$ should be detectable by taking $t$ large, and it would be worthwhile to try to precisely understand this limit the same way $\e^t$ was studied in Example 2.

\section{Concluding remarks}

To summarize, this note made two main points:
\begin{enumerate}
  \item Any physical theory must be finitary if it is based on classical mathematical foundations, as a rigorous definition ultimately regularizes all infinities.
  \item Many regulators that are needed to rigorously define modern QFTs are never discussed. They are not unphysical, and one can look for ways to experimentally probe them.
\end{enumerate}
These lessons are very broad. They motivate a deeper look at how existing continuum QFTs can arise out of lattice ones \cite{Radicevic:1D, Radicevic:2D, Radicevic:3D}. The examples given here will hopefully inspire the reader to think of even more\ldots\! creative applications, such as detecting corrections to the number $\e$.

A deficiency of space and time precludes this from becoming a more detailed study of the finitist/formalist tension in modern high energy physics. Still, one reference deserves a shout-out. In 2011, Tong presented what is possibly the most cogent ``mainstream'' take on this issue \cite{Tong:2011}. He stressed that in quantum theory it is the discrete (e.g.\ quantized energy levels) that seems to emerge from the continuous (e.g.\ Schr\"odinger equations on smooth manifolds). One point of this paper is that this emergence, while certainly real, \emph{cannot be the whole story}. There is no way to define a differential equation on a smooth manifold without eventually introducing a regularization using another finite (and hence discrete) structure. In loftier words, this paper says that there are \emph{fewer} things in heaven and earth than are dreamt of in your philosophy.

This paper has also avoided discussing any specific discrete models for a fundamental physical theory. Whether our world is made of  cellular automata or matrix models or lattice qubits is besides the point. The goal here is merely to stress that the world cannot be described by formal theories ``all the way down.'' Discussing and discovering the true discrete degrees of freedom remains a task for the far future. With luck, this task will at no point need to devolve into a discussion about ontology in quantum theory.

A discussion that \emph{may} be worth having concerns alternative mathematical takes on the notion of finitism. The analysis here did not require defining ultrafinitism, discussing nonstandard analysis, or distinguishing between, say, constructivism and intuitionism. It would also be intriguing if there were ways to escape the finitist axioms by moving away from classical logic.

The ideas presented here might never take hold in the community. Conversely, decades from now, the insistence of today's physicists on formal theories might be judged to have been as silly as their forebearers' view that nature is governed by classical mechanics. Only time will tell.

\section*{Acknowledgments}

This work was completed with the support from the Simons Foundation through \emph{It from Qubit: Simons Collaboration on Quantum Fields, Gravity, and Information}, and from the Department of Energy Office of High-Energy Physics grant DE-SC0009987 and QuantISED grant DE-SC0020194.

\newpage

\bibliographystyle{ssg}
\bibliography{Refs}

\begingroup\raggedright\begin{thebibliography}{10}

\bibitem{Kramer:1983}
E.~E. Kramer, {\em The Nature and Growth of Modern Mathematics}.
\newblock PU Press, 1983.

\bibitem{Brouwer:1908}
L.~E.~J. Brouwer, {\em Die onbetrouwbaarheid der logische principes}.
\newblock P. Noordhoff, 1908.
\newblock Translated, with commentary, in
  \href{https://arxiv.org/abs/1511.01113v1}{\texttt{arXiv:1511.01113}}.

\bibitem{Heijenoort:1967}
J.~van Heijenoort, {\em A Source Book in Mathematical Logic, 1879-1931}.
\newblock Harvard University Press, 1967.
\newblock This book contains papers by Brouwer, Hilbert, Kolmogorov, and Weyl
  that set up the analogue of the finitist/formalist rift in mathematical
  logic.

\bibitem{Snapper:1979}
E.~Snapper, ``The Three Crises in Mathematics: Logicism, Intuitionism and
  Formalism,'' {\em Mathematics Magazine} {\bf 52} (1979), no.~4 207--216.

\bibitem{Bishop:1985}
E.~Bishop, ``Schizophrenia in contemporary mathematics,'' in {\em Errett
  Bishop: reflections on him and his research}, pp.~1--32, Amer. Math. Soc.,
  1985.

\bibitem{Martin-Lof:2008}
P.~Martin-L{\"o}f, ``The Hilbert-Brouwer controversy resolved?,'' in {\em One
  Hundred Years of Intuitionism (1907--2007): The Cerisy Conference} (M.~van
  Atten, P.~Boldini, M.~Bourdeau, and G.~Heinzmann, eds.), pp.~243--256,
  Birkh{\"a}user Basel, 2008.

\bibitem{Lenchner:2020}
J.~Lenchner, ``A Finitist's Manifesto: Do we need to Reformulate the
  Foundations of Mathematics?,'' \href{https://arxiv.org/abs/2009.06485}{{\tt
  2009.06485}}.

\bibitem{Bousso:2002ju}
R.~Bousso, ``{The Holographic principle},'' {\em Rev. Mod. Phys.} {\bf 74}
  (2002) 825--874, \href{https://arxiv.org/abs/hep-th/0203101}{{\tt
  hep-th/0203101}}.

\bibitem{Wilson:1971bg}
K.~G. Wilson, ``{Renormalization group and critical phenomena. 1.
  Renormalization group and the Kadanoff scaling picture},'' {\em Phys. Rev.}
  {\bf B4} (1971) 3174--3183.

\bibitem{Gelfand:1964}
I.~M. Gel'fand and N.~J. Vilenkin, {\em Generalized Functions, vol. 4: Some
  Applications of Harmonic Analysis. Rigged Hilbert Spaces}.
\newblock Academic Press, New York, 1964.

\bibitem{Reed:1972}
M.~Reed and B.~Simon, {\em Methods of modern mathematical physics I: Functional
  analysis}.
\newblock Academic Press, New York, 1972.

\bibitem{tHooft:1973alw}
G.~'t~Hooft, ``{A Planar Theory for Strong Interactions},'' {\em Nucl. Phys. B}
  {\bf 72} (1974) 461.

\bibitem{Gisin:2018ihz}
N.~Gisin, ``{Indeterminism in Physics, Classical Chaos and Bohmian Mechanics.
  Are Real Numbers Really Real?},'' {\em Erkenntnis} {\bf 86} (2021)
  1469--1481, \href{https://arxiv.org/abs/1803.06824}{{\tt 1803.06824}}.

\bibitem{Gisin:2019emq}
N.~Gisin, ``{Real Numbers are the Hidden Variables of Classical Mechanics},''
  {\em Quant. Stud. Math. Found.} {\bf 7} (2019), no.~2 197--201,
  \href{https://arxiv.org/abs/1909.04514}{{\tt 1909.04514}}.

\bibitem{Radicevic:2D}
{\DJ}.~Radi{\v c}evi{\'c}, ``{The UV Structure of QFTs. Part 2: What is Quantum
  Field Theory?},'' \href{https://arxiv.org/abs/2105.12147}{{\tt 2105.12147}}.

\bibitem{Radicevic:1D}
{\DJ}.~Radi{\v c}evi{\'c}, ``{The UV Structure of QFTs. Part 1: Quantum
  Mechanics},'' \href{https://arxiv.org/abs/2105.11470}{{\tt 2105.11470}}.

\bibitem{Radicevic:3D}
{\DJ}.~Radi{\v c}evi{\'c}, ``{The UV Structure of QFTs. Part 3: Gauge
  Theories},'' \href{https://arxiv.org/abs/2105.12751}{{\tt 2105.12751}}.

\bibitem{Tong:2011}
D.~Tong, ``Physics and the integers.'' Available at
  \href{https://www.damtp.cam.ac.uk/user/tong/talks/integer.pdf}{\texttt{www.damtp.cam.ac.uk/user/tong/talks/integer.pdf}},
  2011.

\end{thebibliography}\endgroup

\end{document}